\documentstyle[preprint,aps,floats,epsfig]{revtex}

\begin{document}

\draft

\preprint{\rightline{ANL-HEP-PR-97-28}}

\title{Thermodynamics of Lattice QCD with Chiral 4-Fermion Interactions}

\author{J.~B.~Kogut}
\address{Department of Physics, University of Illinois, 1110 West Green Street,
Urbana, IL 61801, USA}
\author{J.-F.~Laga\"{e} and D.~K.~Sinclair}
\address{HEP Division, Argonne National Laboratory, 9700 South Cass Avenue,
Argonne, IL 60439, USA}

\maketitle

\begin{abstract}
We have studied lattice QCD with an additional, irrelevant 4-fermion
interaction having a $U(1) \times U(1)$ chiral symmetry, at finite temperatures.
Adding this 4-fermion term allowed us to work at zero quark mass, which would
have otherwise been impossible. The theory with 2 massless staggered quark
flavours appears to have a first order finite temperature phase transition at
$N_t=4$ for the value of 4-fermion coupling we have chosen, in contrast to
what is expected for 2-flavour QCD. The pion screening mass is seen to vanish
below this transition, only to become massive and degenerate with the $\sigma$
($f_0$) above this transition where the chiral symmetry is restored, as is seen
by the vanishing of the chiral condensate. 
%$f_0$ --- $a_0$ degeneracy does not appear
%to be restored above this transition, indicating that the flavour singlet axial
%symmetry $U(1)_{axial}$ is not yet restored. 
\end{abstract}


\setcounter{page}{1}
\pagestyle{plain}
\parskip 5pt
\parindent 0.5in

\section{INTRODUCTION}
Studying the finite temperature phase transition of lattice QCD and the
equation of state near this transition requires an understanding of the zero
quark mass limit, where molecular dynamics methods fail completely
\cite{karsch4,milc6}. Even at realistic values of the $u$ and $d$ quark masses,
the Dirac operator is nearly singular, and iterative methods for its inversion
become extremely costly in computer time. \footnote {For the status of lattice
QCD thermodynamics with the standard staggered action and two light flavours we
refer the reader to recent publications \cite{milc12,htmcgc8}. Earlier work is
summarised and referenced in recent reviews \cite{ukawa96,kanaya95,karsch93}.}
For this reason, we modify the lattice QCD action by the addition of an
irrelevant, chirally invariant 4-fermion interaction which renders the Dirac
operator non-singular, even when the quark mass is zero. Because the extra
interaction is irrelevant, such an action should lie in the same universality
class as the standard action, and thus have the same continuum limit. The
4-fermion interaction we choose is of the Gross-Neveu, Nambu-Jona-Lasinio form
\cite{gn,njl}. Ideally, such an interaction should be chosen to have the
$SU(N_f) \times SU(N_f)$ flavour symmetry of the original QCD action. However,
we note that when one introduces auxiliary scalar and pseudoscalar fields to
render this action quadratic in the fermion fields --- which is necessary for
lattice simulations, --- the fermion determinant is no longer real, even in the
continuum limit. Thus for 2 flavour QCD ($N_f=2$), we make a simpler choice and
choose a 4-fermion term with the symmetry $U(1) \times U(1) \subset SU(2)
\times SU(2)$, where $U(1) \times U(1)$ is generated by
$(\tau_3,\gamma_5\tau_3)$. The euclidean Lagrangian density for this theory is
then 
\begin{equation}
{\cal L}=\frac{1}{4}F_{\mu\nu}F_{\mu\nu}
        +\bar{\psi}(D\!\!\!\!/+m)\psi
        -{\lambda^2 \over 6 N_f}[(\bar{\psi}\psi)^2
                          -(\bar{\psi}\gamma_5\tau_3\psi)^2].
\label{eqn:lagrangian}
\end{equation}

Lattice field theories incorporating fermions interacting both through gauge
fields and through quartic self-interactions have been studied before --- see
for example \cite{kmy}. Brower et al. \cite{brower} have suggested the addition
of such chiral 4-fermion interactions to lattice QCD to control the singular
nature of the zero mass Dirac operator. In addition, 4-fermion terms arise in
systematic attempts to improve the fermion lattice action to make it better
approximate the continuum action \cite{sw,milc}. Our work was suggested by
earlier work by one of us on lattice field theories with quartic 4-fermion
actions \cite{hands1,kim,hands2} and by studies of the role such terms play in
lattice QED. 

We have simulated this theory using 2 flavours of staggered quarks on $8^3
\times 4$ and $12^2 \times 24 \times 4$ lattices, at an intermediate value of
$\lambda$ and zero quark mass, in order to determine the position and nature of
the finite temperature transition. We also present some zero temperature
results on an $8^3 \times 24$ lattice, where we demonstrate that the theory
with massless quarks does indeed have a massless Goldstone pion. In addition to
measuring the standard order parameters we have measured the pion,
$\sigma$($f_0$), and $a_0$ screening masses to probe the nature of chiral
symmetry restoration at this transition. We also simulated the corresponding
theory with 4-fermion couplings but no gauge interactions on relatively small
lattices ($8^4$ and $8^3 \times 4$) to aid us in deciding what values of
4-fermion coupling constant to choose. 

In section 2 we discuss the lattice formulation of QCD with chiral 4-fermion
interactions. We present our zero gauge-coupling results in section 3. The zero
temperature results are given in section 4, while the finite temperature
simulations and results are described in section 5. Section 6 gives discussions
and conclusions, and outlines directions for future research. 
 
\section{Lattice QCD with chiral 4-fermion interactions}

Equation~\ref{eqn:lagrangian} can be rendered quadratic in the fermion fields
by the standard trick of introducing (non-dynamical) auxiliary fields $\sigma$ 
and $\pi$ in terms of which this Lagrangian density becomes
\begin{equation}
{\cal L}=\frac{1}{4}F_{\mu\nu}F_{\mu\nu}
        +\bar{\psi}(D\!\!\!\!/+\sigma+i\pi\gamma_5\tau_3+m)\psi
        +{3 N_f \over 2 \lambda^2}(\sigma^2+\pi^2)
\end{equation}
The molecular dynamics Lagrangian for a particular staggered fermion lattice
transcription of this theory in which $\tau_3$ is identified with $\xi_5$,
the flavour equivalent of $\gamma_5$ is
\begin{eqnarray}
L & = & -\beta\sum_{\Box}[1-\frac{1}{3}{\rm Re}({\rm Tr}_{\Box} UUUU)]
        +{N_f \over 8}\sum_s \dot{\psi}^{\dag} A^{\dag} A\dot{\psi}
        -\sum_{\tilde{s}}\frac{1}{8}N_f\gamma(\sigma^2+\pi^2)  \nonumber \\
  &   & +\frac{1}{2}\sum_l(\dot{\theta}_7^2+\dot{\theta}_8^2
        +\dot{\theta}_1^{\ast}\dot{\theta}_1
        +\dot{\theta}_2^{\ast}\dot{\theta}_2
        +\dot{\theta}_3^{\ast}\dot{\theta}_3)
        +\frac{1}{2}\sum_{\tilde{s}}(\dot{\sigma}^2+\dot{\pi}^2)
\end{eqnarray}
where
\begin{equation}
A = \not\!\! D + m + \frac{1}{16} \sum_i (\sigma_i+i\epsilon\pi_i)
\end{equation}
with $i$ running over the 16 sites on the dual lattice neighbouring the site
on the normal lattice, $\epsilon=(-1)^{x+y+z+t}$ and $\not\!\! D$ the usual
gauge-covariant ``d-slash'' for the staggered quarks. The factor ${N_f \over 8}$
in front of the pseudo-fermion kinetic term is appropriate for the hybrid
molecular dynamics algorithm with ``noisy'' fermions, where $A\dot{\psi}$ are
chosen from a complex gaussian distribution with width 1. The ``dots''
represent derivatives with respect to molecular dynamics ``time'' as distinct
from normal time. For the presentation of all our simulation results we use
a time definition which is twice this, in order to be consistent with the
convention used in the works of the HEMCGC and HTMCGC collaborations. We note
that $\gamma=3/\lambda^2$. Although the determinant of $A$ does not appear to
be real, it becomes so in the continuum limit. Without the gauge fields, this
theory reverts to the one studied in \cite{kim}, with $3 N_f$ flavours.

The advantage of this choice of the chiral 4-fermion interaction is that it 
preserves the axial $U(1)$ chiral symmetry of the normal staggered quark
lattice QCD action generated by $\gamma_5\xi_5$ at $m=0$. This means that,
when chiral symmetry is spontaneously broken, the pion associated with
$\xi_5\gamma_5$ will be a true Goldstone boson and will be massless at $m=0$,
even for finite lattice spacing. Under this exact chiral symmetry the fields
transform as
\begin{eqnarray}
\dot{\psi}(n) & \rightarrow & e^{-i\frac{1}{2}\phi\epsilon(n)}\dot{\psi}(n) \\
\sigma(n)+i\pi(n) & \rightarrow & e^{i\phi}[\sigma(n)+i\pi(n)]
\label{eqn:chiral}
\end{eqnarray}
from which we find that
\begin{eqnarray}
A\dot{\psi}(n) & \rightarrow & e^{i\frac{1}{2}\phi\epsilon(n)}A\dot{\psi}(n) \\
\sigma(n)+i\epsilon(n)\pi(n) & \rightarrow & e^{i\phi\epsilon(n)}
                                               [\sigma(n)+i\epsilon(n)\pi(n)],
\end{eqnarray}
when $m=0$. Hence, for massless quarks the above Lagrangian has an exact $U(1)$
flavour axial symmetry.

\section{The Nambu-Jona-Lasinio-Gross-Neveu model.}

In order to ascertain which ranges of values of $\gamma$ represent strong
coupling and which represent weak coupling, we simulated the above lattice
theory without the gluon fields where it becomes the lattice version of the
Nambu-Jona-Lasinio-Gross-Neveu model with $3 N_f = 6$ flavours
\cite{njl,gn,kim}. We have simulated this theory with $m=0$ on $8^4$ and $8^3
\times 4$ lattices. For these simulations, we ran for 5000 molecular-dynamics
time units at each $\gamma$ value with step size $dt=0.05$. 

\begin{figure}[htb] 
\epsfxsize=4in
\centerline{\epsffile{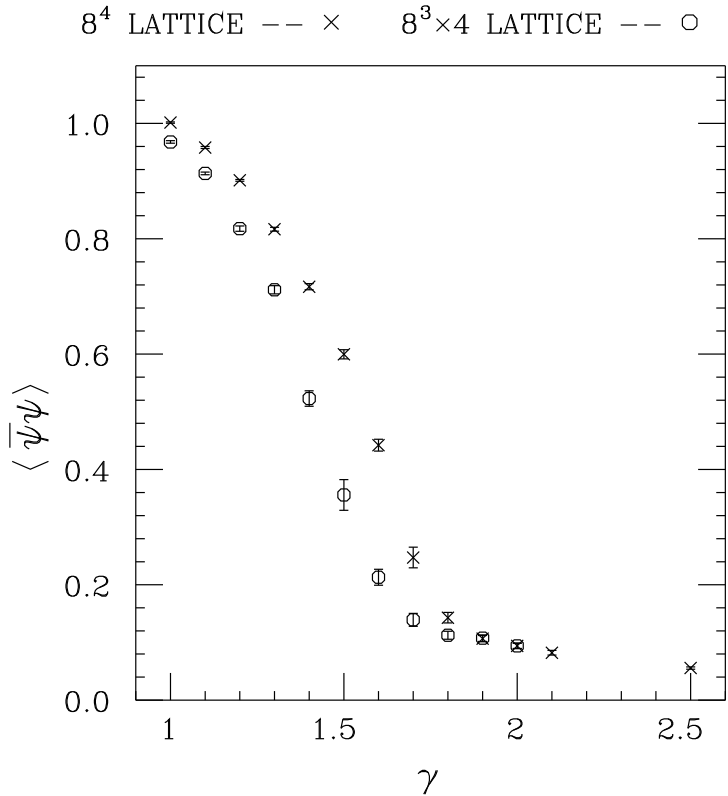}}
\caption{$\langle\bar{\psi}\psi\rangle$ as a function of $\gamma$ for the pure
4-fermion theory for both zero and finite temperatures.\label{fig:njlgn}}
\end{figure}

This theory is known to have a phase transition, even at zero temperature, as a
function of $\gamma$ \cite{hkk}. For strong coupling (small $\gamma$) chiral
symmetry is spontaneously broken, while at weak coupling (large $\gamma$)
chiral symmetry remains unbroken. Since the theory is strictly massless, the
direction this axial symmetry is broken is arbitrary, and the chiral condensate
is a linear combination of $\langle\bar{\psi}\psi\rangle$ and
$i\langle\bar{\psi}\gamma_5\xi_5\psi\rangle$ (or $\sigma$ and $\pi$). Because
we work on finite lattices, the direction of this symmetry breaking does not
remain constant, but rotates over the course of the run. For this reason, the
order parameter we will denote by $\langle\bar{\psi}\psi\rangle$ is actually
the molecular-dynamics time average of $\sqrt{\langle\bar{\psi}\psi\rangle^2
-\langle\bar{\psi}\gamma_5\xi_5\psi\rangle^2}$, which differs from the true
chiral condensate by terms ${\cal O}(1/\sqrt{V})$ where $V$ is the space-time
volume of the lattice. This quantity is given in figure~\ref{fig:njlgn} as a
function of $\gamma$ for both lattice sizes. 

From the $8^4$ lattice we conclude that the zero temperature phase transition
from the strong coupling phase to the weak coupling phase occurs somewhere in
the range $1.7 \lesssim \gamma \lesssim 1.8$, while the finite temperature
transition on the $8^3 \time 4$ lattice occurs at a somewhat smaller value of
$\gamma$. Thus we would conclude that since, for lattice QCD with chiral
4-fermion interactions, we wish to work in the weak 4-fermion coupling regime,
we should restrict ourselves to values of $\gamma$ which are greater than 2.
Moreover, since the introduction of this 4-fermion term produces chiral phase
transitions where none previously existed, we could expect that chiral symmetry
might be restored at a temperature higher than the deconfinement temperature.
Since this is not the case for normal lattice QCD, we should try to work at
a large enough $\gamma$ value that the 2 transitions are close, if not 
coincident. We describe studies to determine the appropriate range of $\gamma$
values in section~5.

\section{Zero Temperature Results}

We have undertaken a preliminary investigation of the hadron spectrum of QCD
with chiral 4-fermion interactions of the type discussed in section~2 on an
$8^3 \times 24$ lattice with 2 flavours of massless quarks at $\beta=6/g^2=5.4$
and $\gamma=10$. $\beta=5.4$ was chosen since it lies just below the phase
transition for lattice QCD without the 4-fermion term on an $N_t=8$ lattice.
Thus we were assured of being in the confined phase. We used the hybrid
molecular dynamics method with time increment $dt=0.05$ for updating. We ran
for a total of 5000 time units, discarding the first 1000 time units for
equilibration. 250 configurations spaced by 20 time units were saved for
further analysis. The first 50 of these were discarded for equilibration. 

The pion and $\sigma$ propagators were calculated from the correlations of the
$\pi$ and $\sigma$ fields, sampled every 2 time units. Because, as was the
case in the previous section, the direction of chiral symmetry breaking rotates
during a run on a finite lattice, we chirally rotated each $(\sigma,\pi)$
configuration (see equation~\ref{eqn:chiral}) so that $\sum_{sites}\pi = 0$.
The propagators are then given by
\begin{eqnarray}
P_{\sigma}(T) &=& {1 \over V} \sum_t\langle\sum_{\bf x}\sigma({\bf x},t)
                              \sum_{\bf y}\sigma({\bf y},t+T)\rangle 
                              -v \langle\sigma\rangle^2
\label{eqn:sigma}                                                      \\
P_{\pi}(T)     &=& {1 \over V} \sum_t\langle\sum_{\bf x}\pi({\bf x},t)
                              \sum_{\bf y}\pi({\bf y},t+T)\rangle 
\end{eqnarray}
where $V$ is the space-time and $v$ is the spatial volume of the lattice.
Correlated fits are made to these propagators, binning over 10 measurements
(20 time units) to account for correlations, and using jackknife to remove
the vacuum expectation values in the $\sigma$ propagator. 

\begin{figure}[htb] 
\epsfxsize=4in
\centerline{\epsffile{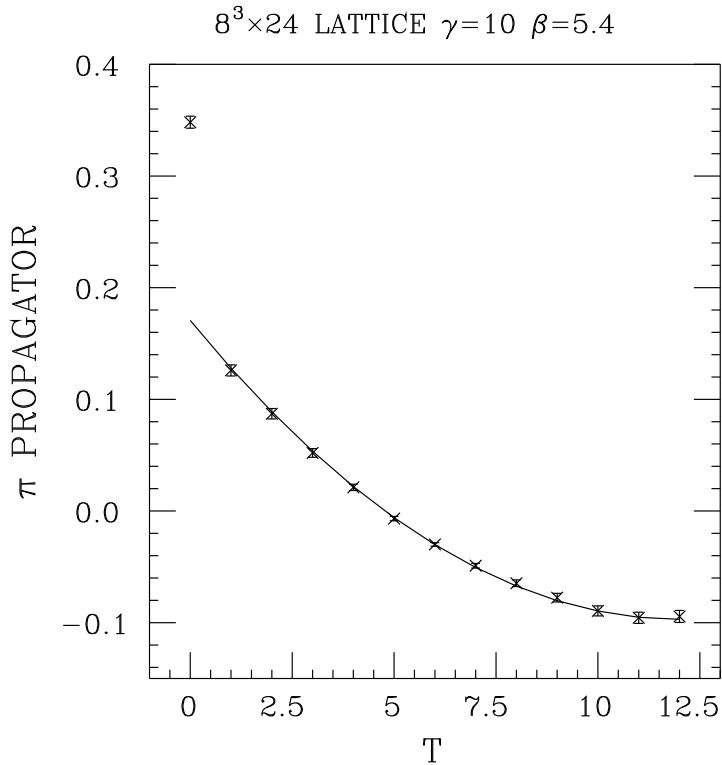}}
\caption{The $\pi$ propagator at zero temperature, the curve is a fit to a
massless scalar boson.\label{fig:pi}}
\end{figure}

The $\pi$ propagator was fitted to the form
\begin{equation}
P_{\pi}(T) = A P_0(T) - B
\end{equation}
where $P_0(T)$ is the lattice propagator for a massless scalar boson
\begin{equation}
P_0(T)=\frac{1}{2N_t}\sum_{k=-N_t/2+1}^{N_t/2}{e^{2\pi i k T/N_t} \over
                                                     (1-cos(2\pi k/N_t)},
\end{equation}          
with the $k=0$ mode excluded. We found $A=3.7(1)$ and $B=7.5(1)$ for the fit
from $T=1$ to $T=12$. This fit has a 78\% confidence level. Figure~\ref{fig:pi}
shows the measured propagator with this fit superimposed. As indicated by the
high confidence level, our measured pion propagator is very well fit by a
massless scalar propagator, except at $T=0$ where we see the remnant of the
original $\delta$-function interaction. Thus the sum of fermion ``bubble''
diagrams has converted the auxilliary $\pi$ field to a true Goldstone pion.

\begin{figure}[htb] 
\epsfxsize=4in
\centerline{\epsffile{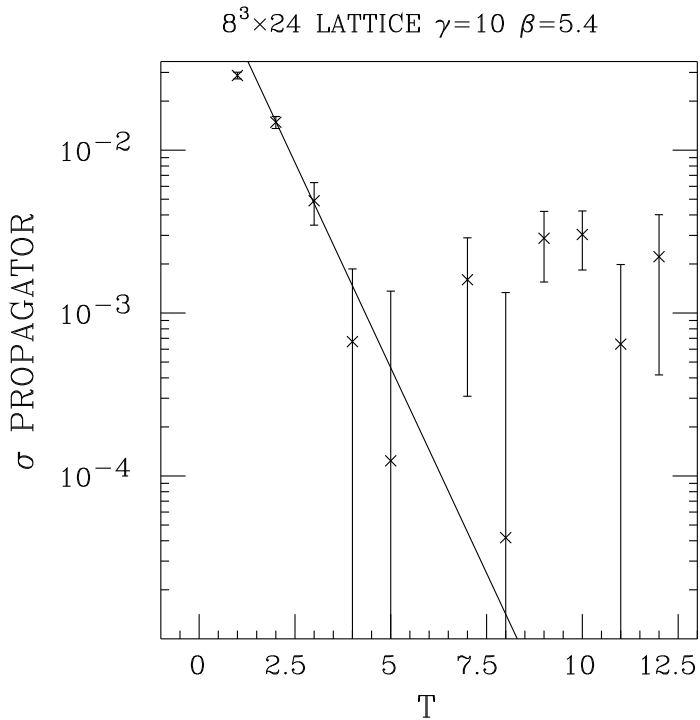}}
\caption{The $\sigma$ propagator at zero temperature, the curve is the single
particle fit described in the text.\label{fig:sigma}}
\end{figure}

For large $T$ the sigma propagator should behave as
\begin{equation}
P_{\sigma}(T) = A [e^{-m_\sigma T} + e^{-m_\sigma (N_t-T)}]
              + B [e^{-m_{\pi_2} T} + e^{-m_{\pi_2} (N_t-T)}]
\end{equation}
We have attempted such fits, but are unable to find a stable value for the
$\pi_2$ mass. For this reason we resorted to the single particle fit obtained
by setting $B$ to zero. Here, our best fit yielded $m_\sigma = 1.16(24)$ with
a 51\% confidence level. This mass was consistent with those obtained from
the 2 particle fits. We have plotted our $\sigma$ propagator and this fit in
figure~\ref{fig:sigma}. Here it is clear why our fits were so poor --- the
propagator disolves into noise after only a few time-slices, which is the
usual problem encountered when trying to calculate high mass propagators as
in equation~\ref{eqn:sigma}. (Here the error in the propagator is expected to
be independent of T, whereas for meson propagators calculated directly from the
quark propagator, the error falls with increasing T). It is interesting to
speculate as to whether the ``tail'' of this propagator is due to the 2-pion
cut in the $\sigma$ propagator. 

\section{Finite temperature results}

\begin{figure}
\centerline{\hbox{\psfig{figure=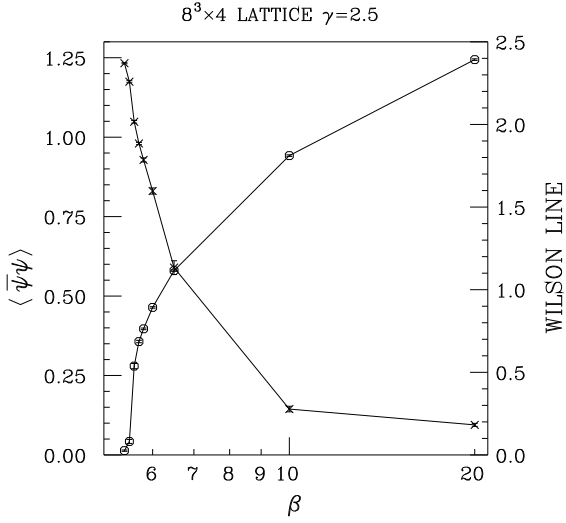,height=8cm,width=8cm},
                  \psfig{figure=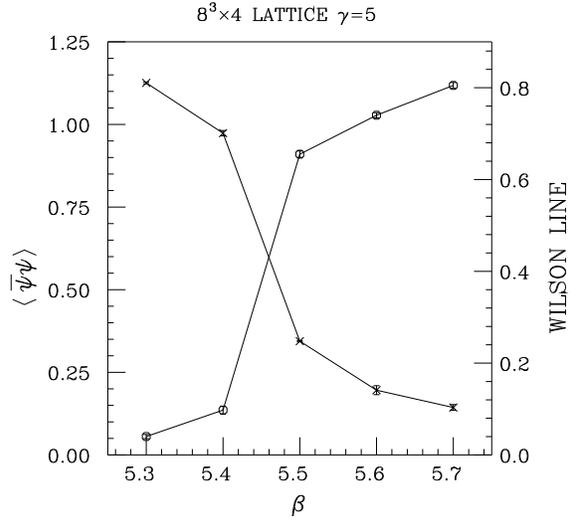,height=8cm,width=8cm}}}
\centerline{(a) \hspace{2.5in} (b)}
\vspace{0.1in}
\centerline{\hbox{\psfig{figure=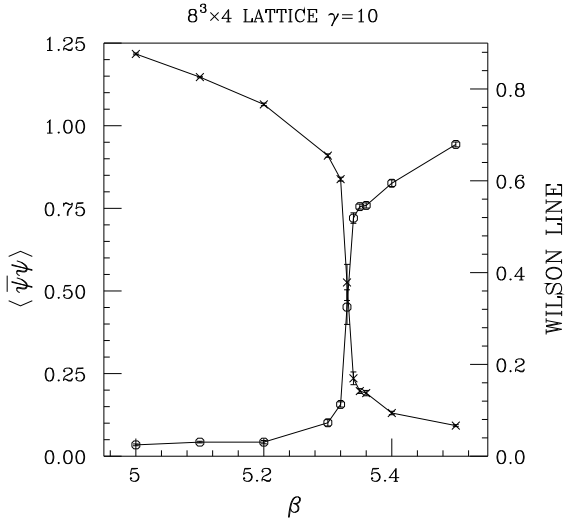,height=8cm,width=8cm},
                  \psfig{figure=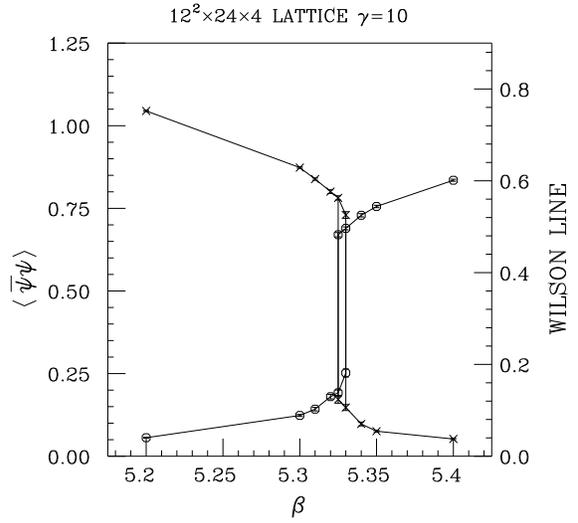,height=8cm,width=8cm}}}
\centerline{(c) \hspace{2.5in} (d)}
\vspace{0.1in}
\caption{Wilson/Polyakov line (circles) and $\langle\bar{\psi}\psi\rangle$ 
         (crosses) versus $\beta$ for a) An $8^3 \times 4$ lattice at 
         $\gamma=2.5$, b) An $8^3 \times 4$ lattice at $\gamma=5$, 
         c) An $8^3 \times 4$ lattice at $\gamma=10$, and 
         d) A $12^2 \times 24 \times 4$ lattice at $\gamma=10$.
         \label{fig:wil-psi}}
\end{figure}

We have performed simulations on $N_t=4$ lattices and with 2 flavours of zero
mass quarks, in order to study the transition from hadronic matter to a quark
gluon plasma at finite temperature. Our initial runs were performed on $8^3
\times 4$ lattices with $\gamma=2.5$, $5$, and $10$ The time step $dt=0.05$ for
these runs. We experimented with values of $dt$ from $0.02$ to $0.20$, and
found $dt=0.05$ to be small enough that finite $dt$ errors were smaller than
statistical errors. We measured the Wilson/Polyakov line and the chiral
condensate, $\langle\bar{\psi}\psi\rangle$, enabling us to determine the phase
structure of the theory. Since, as we have seen in section~3, the presence of
the 4-fermion coupling can break chiral symmetry, even at zero gauge coupling
where there is no confinement, we expect that the chiral symmetry restoration
and deconfinement will in general occur at different values of $\beta$. In fact
for values of $\gamma$ less than that for the transition at zero gauge
coupling, chiral symmetry will always be broken. In figure~\ref{fig:wil-psi} we
show the values of the 2 order parameters for each of $\gamma=2.5$, $5$, and
$10$. It is clear at $\gamma=2.5$ that the deconfinement transition, marked by
the rapid increase in the Wilson/Polyakov line (around $\beta=5$), occurs at 
a much lower value of
$\beta$ than the chiral symmetry restoration, marked by a drop in
$\langle\bar{\psi}\psi\rangle$ (which does not reach
 zero for the reasons given in section~3) between $\beta=6$ and $10$. 
At $\gamma=5$ the 2 transitions still appear to be
distinct, although we cannot rule out this being a finite volume effect, while
at $\gamma=10$ the 2 transitions appear to be coincident. Since we are
interested in obtaining results which are relevant to the continuum limit where
the 4-fermion coupling vanishes and the 2 transitions {\it are} believed to be
coincident, we chose to work at $\gamma=10$. 

\begin{figure}[htb] 
\epsfxsize=4in
\centerline{\epsffile{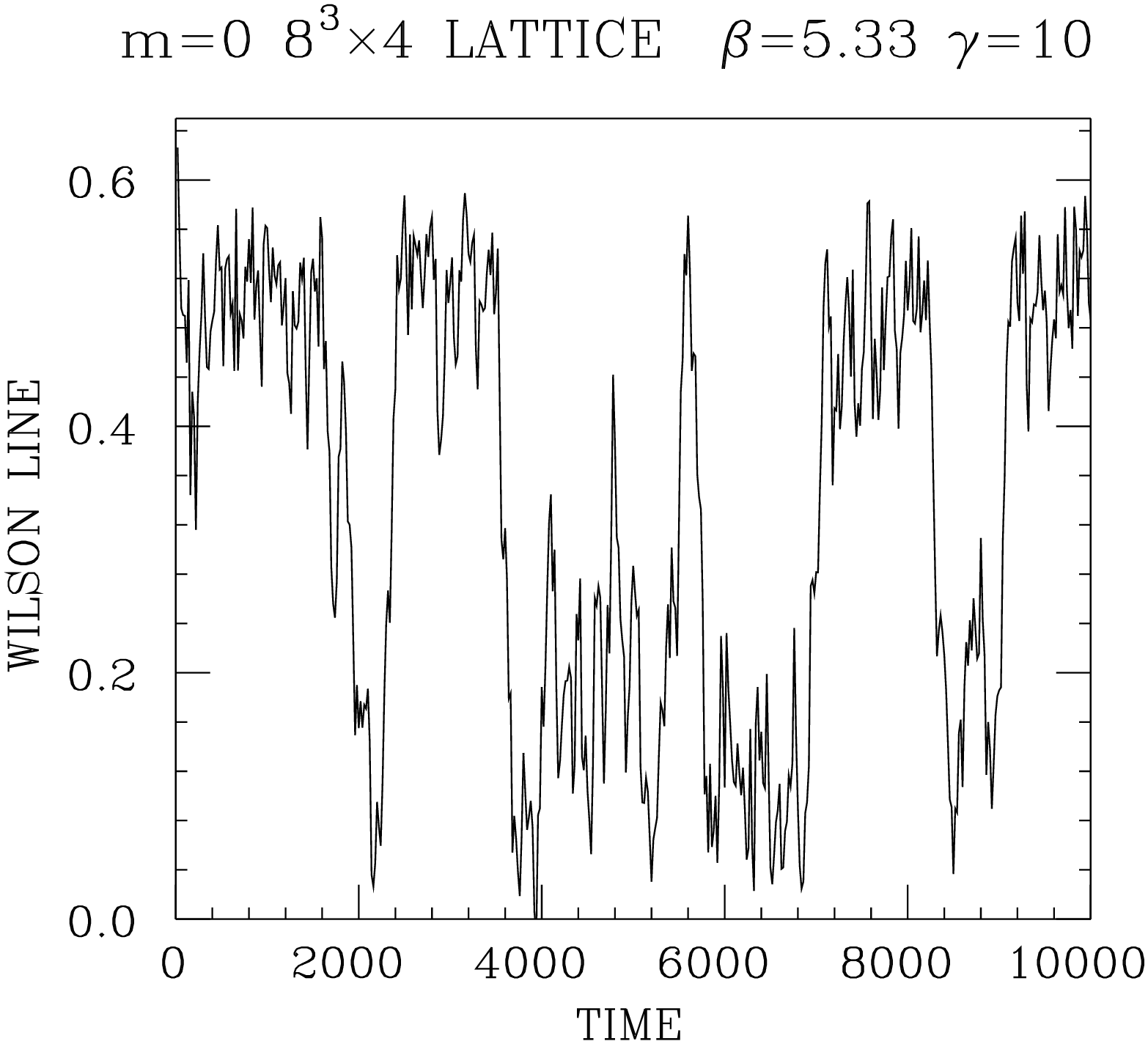}}
\caption{The time evolution of the Wilson/Polyakov line at $\beta=5.33$ on
an $8^3 \times 4$ lattice.\label{fig:533_8}}
\end{figure}

\begin{figure}
\centerline{\hbox{\psfig{figure=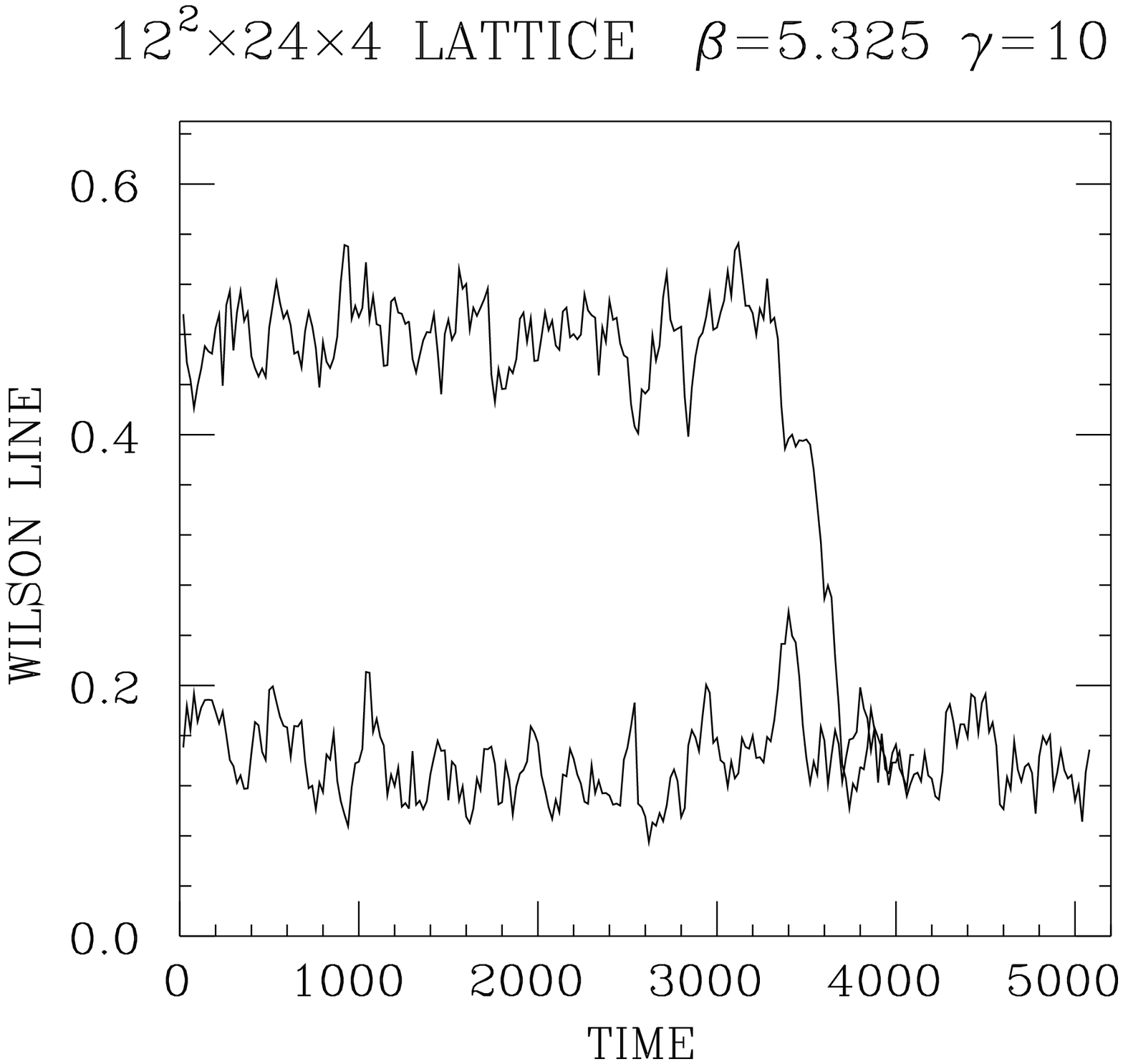,height=8cm,width=8cm},
                  \psfig{figure=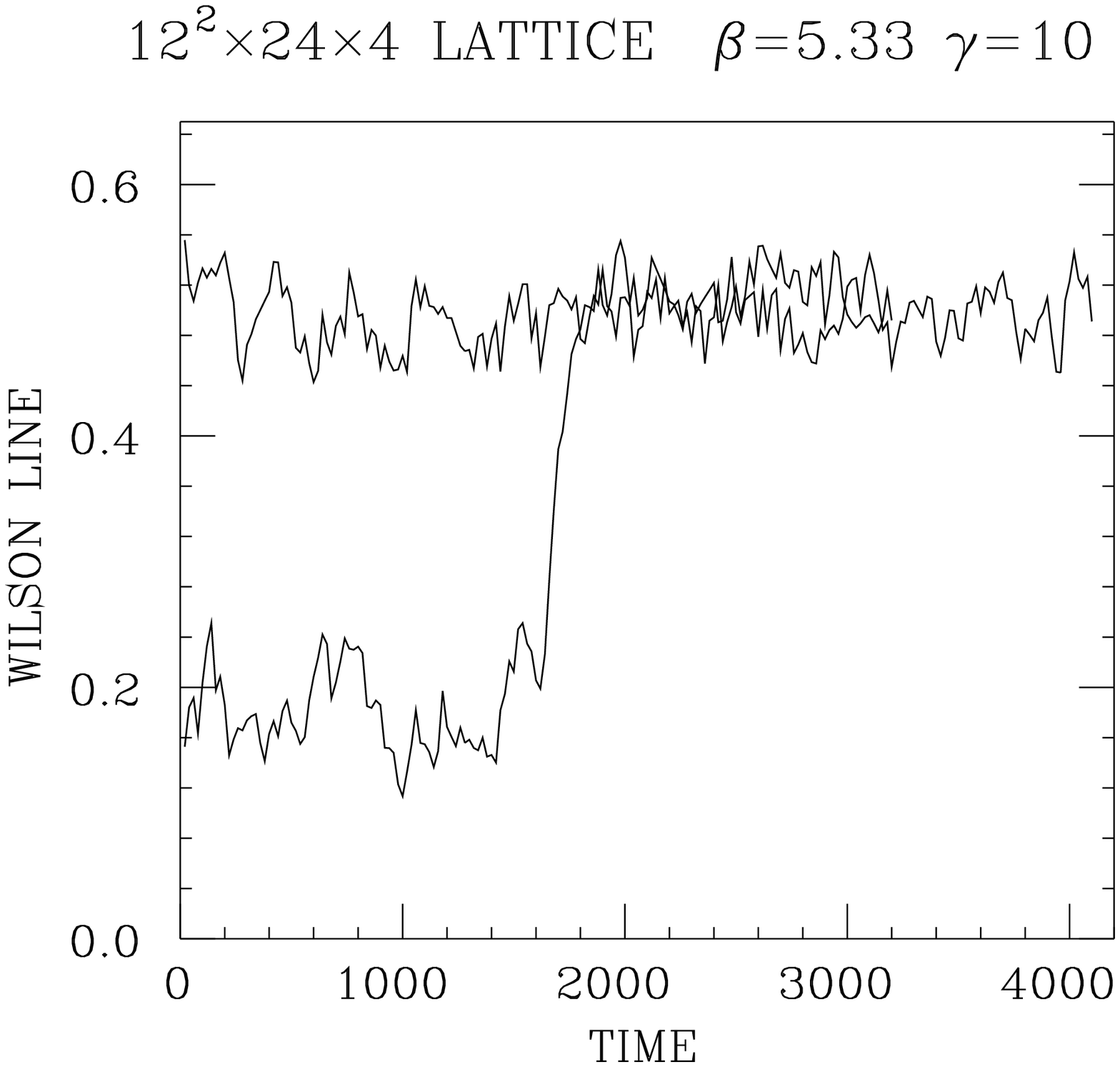,height=8cm,width=8cm}}}
\centerline{(a) \hspace{2.5in} (b)}
\vspace{0.1in}
\caption{Time evolution of the Wilson/Polyakov line on a 
$12^3 \times 24 \times 4$ lattice a) at $\beta=5.325$ and b) at $beta=5.33$.
\label{fig:wilson}}
\end{figure}

At $\gamma=10$, we observed signs of a two state signal in the time evolution
of both order parameters at $\beta=5.33$, indicating the possibility of a first
order transition. Figure~\ref{fig:533_8} shows the time evolution of the
Wilson/Polyakov line at this $\beta$ value. Thus we conclude that the transition
occurs at $\beta=5.330(5)$ on this size lattice. Since the standard action 
shows a false first order transition on small lattices, we repeated our runs on
a $12^2 \times 24 \times 4$ lattice. Having $N_z=24$ enabled us to measure
hadronic screening lengths in the $z$ direction.
On this larger lattice, we find evidence for 2-state signals at $\beta=5.325$
and $\beta=5.33$. In figure~\ref{fig:wilson} we show the time evolution of the
Wilson/Polyakov line for these 2 $\beta$ values. At $\beta=5.325$ the Wilson
line measured from a cold start remains small for the total length of the
run while that from a hot start, remains large for over 3000 time units before
it tunnels rapidly to a low value and stays there. At $\beta=5.33$ the Wilson
line from a hot start remains high for the entire run, while that from a cold
start remains low for over 1500 time units before it tunnels rapidly to a high
value and stays there. There is no sign of metastability for higher or lower
$\beta$ values. We thus conclude that the transition occurs in the range
$5.325 \lesssim \beta \lesssim 5.33$, and probably closer to $\beta=5.325$.
The values of these quantities, $\langle\sigma\rangle$ and the
average plaquette, $\langle 1 - \frac{1}{3} {\rm Tr} U_\Box \rangle$ are
given in tables~\ref{tab:wilpsiapq8},\ref{tab:wilpsiapq12}. We note that the
required relationship 
\begin{equation} 
\langle\bar{\psi}\psi\rangle = \gamma \langle\sigma\rangle
\end{equation}
is approximately true in the chirally broken phase, but that in the chirally
symmetric phase it is somewhat less well obeyed. Since the continuum values
of both these quantities are zero in this symmetric phase, this suggests that
the difference is in the $\sqrt{1/V}$ departures of our estimators for these
quantities from their true values, which do not have to obey this relationship.

\begin{table}[htb]
\begin{tabular}{lldddd}
$\beta$ & time  & Wilson Line & $\langle\bar{\psi}\psi\rangle$ 
& $\langle\sigma\rangle$ & PLAQUETTE  \\
\hline
5.0     & 2000  &  0.025(2)   & 1.217(3)  & 0.1235(5)  & 0.5825(3)  \\
5.1     & 2000  &  0.031(2)   & 1.147(2)  & 0.1170(5)  & 0.5636(3)  \\
5.2     & 1000  &  0.031(4)   & 1.065(4)  & 0.1080(7)  & 0.5430(8)  \\
5.3     & 1000  &  0.073(7)   & 0.910(5)  & 0.0923(8)  & 0.5149(7)  \\
5.32    & 6000  &  0.113(6)   & 0.838(6)  & 0.0857(8)  & 0.5064(6)  \\
5.33    & 10000 &  0.325(38)  & 0.526(55) & 0.0575(48) & 0.4900(24) \\
5.34    & 4000  &  0.519(11)  & 0.236(19) & 0.0290(17) & 0.4746(5)  \\
5.35    & 2000  &  0.544(5)   & 0.197(7)  & 0.0260(8)  & 0.4717(4)  \\
5.36    & 2000  &  0.546(6)   & 0.191(8)  & 0.0254(9)  & 0.4699(4)  \\
5.4     & 1000  &  0.595(7)   & 0.131(5)  & 0.0203(8)  & 0.4619(4)  \\
5.5     & 1000  &  0.679(5)   & 0.093(3)  & 0.0173(6)  & 0.4469(3)
\end{tabular}
\caption{Wilson line, $\langle\bar{\psi}\psi\rangle$, $\langle\sigma\rangle$
and average plaquette as functions of $\beta$ for an $8^3 \times 4$ lattice at
$\gamma=10$. The ``time'' is the molecular dynamics time for the run. Typically
1/5 of each run was discarded for equilibration.\label{tab:wilpsiapq8}} 
\end{table}
\begin{table}[htb]
\begin{tabular}{lrdddd}
$\beta$ & time  & Wilson Line & $\langle\bar{\psi}\psi\rangle$ 
& $\langle\sigma\rangle$ & PLAQUETTE  \\
\hline
5.2     & 2140  & 0.040(1)    & 1.045(2)  & 0.1056(3)  & 0.5427(2)  \\
5.3     & 2040  & 0.089(20    & 0.873(2)  & 0.0887(3)  & 0.5136(3)  \\
5.31    & 2080  & 0.103(4)    & 0.838(4)  & 0.0849(5)  & 0.5096(3)  \\
5.32    & 4020  & 0.130(5)    & 0.801(6)  & 0.0809(5)  & 0.5055(4)  \\
5.325c  & 4100  & 0.138(6)    & 0.782(6)  & 0.0794(6)  & 0.5034(4)  \\
5.325h  & 5080  & 0.483(5)    & 0.172(11) & 0.0184(12) & 0.4790(3)  \\
5.33c   & 3200  & 0.181(9)    & 0.730(10) & 0.0742(9)  & 0.4994(6)  \\
5.33h   & 4100  & 0.496(3)    & 0.147(9)  & 0.0163(9)  & 0.4775(2)  \\
5.34    & 2080  & 0.525(4)    & 0.098(6)  & 0.0116(6)  & 0.4743(2)  \\
5.35    & 2080  & 0.544(3)    & 0.076(2)  & 0.0096(3)  & 0.4716(1)  \\
5.4     & 2020  & 0.601(2)    & 0.052(2)  & 0.0079(3)  & 0.4621(1)
\end{tabular}
\caption{Wilson line, $\langle\bar{\psi}\psi\rangle$, $\langle\sigma\rangle$
and average plaquette as functions of $\beta$ for an $12^2 \times 24 \times 4$
lattice at $\gamma=10$. The ``time'' is the molecular dynamics time for the
run. Typically 1/5 of each run was discarded for equilibration. h and c refer
to runs from hot and cold starts. Where a tunneling occurs, we only average
over pretunneling data. 
\label{tab:wilpsiapq12}} 
\end{table}

We now turn to a consideration of the hadronic screening lengths which measure
the propagation of exitations with hadronic quantum numbers in hadronic matter 
and the quark-gluon plasma. In particular these measure the manner in which
chiral symmetry is restored as we pass through the transition from hadronic
matter to a quark-gluon plasma. Here we concentrate on the $\pi$ and 
$\sigma(f_0)$ propagators which can be calculated from the $\pi$ and $\sigma$
auxiliary fields, as in section~2, i.e. from
\begin{eqnarray}
P_{\sigma}(Z) &=& {1 \over V} \sum_z\langle\sum_{xyt}\sigma(x,y,z,t)
                              \sum_{x'y't'}\sigma(x',y',z+Z,t')\rangle 
                              -N_x N_y N_t \langle\sigma\rangle^2      \\
P_{\pi}(Z)     &=& {1 \over V} \sum_z\langle\sum_{xyt}\pi(x,y,z,t)
                              \sum_{x'y't'}\pi(x',y',z+Z,t')\rangle 
\end{eqnarray}
In the hadronic phase we fit the pion propagator to 
\begin{equation}
P_{\pi}(Z) = A P_0(Z) - B
\label{eqn:pi}
\end{equation}
where $P_0(Z)$ is the lattice propagator for a massless scalar boson
\begin{equation}
P_0(Z)=\frac{1}{2N_z}\sum_{k=-N_z/2+1}^{N_z/2}{e^{2\pi i k Z/N_z} \over
                                                     (1-cos(2\pi k/N_z)}
\end{equation}          
with the $k=0$ mode excluded. The $\sigma$ propagator was fitted to
\begin{equation}
P_{\sigma}(Z) = A [e^{-m_\sigma Z} + e^{-m_\sigma (N_z-Z)}]
              + B [e^{-m_{\pi_2} Z} + e^{-m_{\pi_2} (N_z-Z)}]
\end{equation}
and to the special case where $B=0$, which was often adequate in this range of
$\beta$ values. Above the transition, in the quark-gluon plasma phase, since 
$\langle\bar{\psi}\psi\rangle=\gamma\langle\sigma\rangle=0$, one cannot
distinguish $\sigma$ and $\pi$ and it makes most sense to consider
\begin{equation}
P_{\sigma\pi} \equiv P_\sigma + P_\pi,
\end{equation}
which we fit to
\begin{equation}
P_{\sigma\pi}(Z) = A [e^{-m_{\sigma\pi} Z} + e^{-m_{\sigma\pi} (N_z-Z)}]
\end{equation}

Since choosing to fit our pion to a massless scalar boson propagator below the
transition, and to a massive propagator above might seem to be forcing the
result we want, we have looked at fits to the pion propagator obtained as in
section~2 by chirally rotating our propagators so that $\langle\pi\rangle=0$,
and then fitting to \ref{eqn:pi}. The results obtained from our best fits are
given in table~\ref{tab:pifits}. As in the zero temperature case we bin our
data in bins of 10, i.e. 20 time units. Note that, for all our mass fits we
have selected our runs from a cold start for $\beta=5.325$ and from a hot start
for $\beta=5.33$, which folds in our knowledge that the transition occurs in
the range $5.325 < \beta < 5.33$ 
\begin{table}[htb]
\begin{tabular}{lcddd}
$\beta$ & range   & A          & B           & confidence \\
\hline
5.2     & 1 -- 11 & 0.0051(2)  &  0.0069(2)  & 0.832      \\
5.3     & 2 -- 12 & 0.0044(3)  &  0.0079(4)  & 0.639      \\
5.31    & 1 -- 12 & 0.0045(2)  &  0.0072(2)  & 0.739      \\
5.32    & 1 -- 11 & 0.0042(1)  &  0.0075(1)  & 0.248      \\
5.325   & 1 -- 11 & 0.0047(2)  &  0.0069(1)  & 0.858      \\
5.33    & 2 -- 5  & 0.0028(1)  &  0.0133(18) & 0.739      \\
5.34    & 2 -- 12 & 0.0020(1)  &  0.0101(4)  & 0.475      \\
5.35    & 4 -- 10 & 0.0010(1)  &  0.0164(12) & 0.642      \\
5.4     & 3 -- 8  & 0.0006(1)  &  0.0155(8)  & 0.567      \\
5.5     & 3 -- 12 & 0.00009(6) & -0.0035(12) & 0.528
\end{tabular}
\caption{Coefficients $A$ and $B$ for the fit of the pion propagator to a
         massless scalar boson (equation~\protect\ref{eqn:pi}).
         \label{tab:pifits}}
\end{table}
We note that the coefficient $A$ which measures the amount of massless scalar
propagator in this fit drops abruptly as we increase $\beta$ through the
phase transition, and continues to drop as $\beta$ is increased. In addition,
the parameters of the fits become more sensitive to the fitting range as
$\beta$ is increased beyond this transition. Thus we feel justified in 
taking the pion screening mass to be zero below the transition, as required by
Goldstone's theorem, while fitting it to a massive propagator above the
transition. 

Above the transition, our definition of $\langle\bar{\psi}\psi\rangle$ should
not vanish, but rather should scale as $1/\sqrt{v}$, and so vanish in the in
the infinite (spatial) volume limit. In table~\ref{tab:pbp} we compare
$\langle\bar{\psi}\psi\rangle$ measured on the $12^2 \times 24 \times 4$ with
$\sqrt{8^3/(12^2 \times 24}) \times \langle\bar{\psi}\psi\rangle$, measured on
the $8^3 \times 4$ lattice, and the corresponding values of
$\langle\sigma\rangle$ for values of $\beta$ above the transition, 

\begin{table}[htb]
\begin{tabular}{ldddd}
$\beta$      &  $\langle\bar{\psi}\psi\rangle_{12^2 \times 24 \times 4}$
     &  $\sqrt{4/27} \times \langle\bar{\psi}\psi\rangle_{8^3 \times 4}$
     &  $\langle\sigma\rangle_{12^2 \times 24 \times 4}$
     &  $\sqrt{4/27} \times \langle\sigma\rangle_{8^3 \times 4}$    \\
\hline
5.33         & 0.147(9)    & 0.202(21)  & 0.0163(9)   & 0.0221(18) \\
5.34         & 0.098(6)    & 0.091(7)   & 0.0116(6)   & 0.0111(7)  \\
5.35         & 0.076(2)    & 0.076(3)   & 0.0096(3)   & 0.0100(3)  \\
5.4          & 0.052(2)    & 0.050(2)   & 0.0079(3)   & 0.0078(3)
\end{tabular}
\caption{Test of $\langle\bar{\psi}\psi\rangle$ scaling in the high temperature
         phase.\label{tab:pbp}}
\end{table}
Except for $\beta=5.33$ which really should not have been included, since the
$8^3 \times 4$ data really represents a mixture of high and low temperature
phases, scaling is true within our statistical errors. This helps justify our
conclusion that chiral symmetry is indeed restored in the high temperature
phase, and our choice of screening propagator fits based on this assumption.
The fact that this scaling is true for both quantities is an indication that
the departures from the relationship between them is an artifact of the
$\sqrt{1/v}$ difference between these quantities and the vacuum condensates
they represent, as suggested above. 
          
\begin{figure}[htb]
\epsfxsize=4in
\centerline{\epsffile{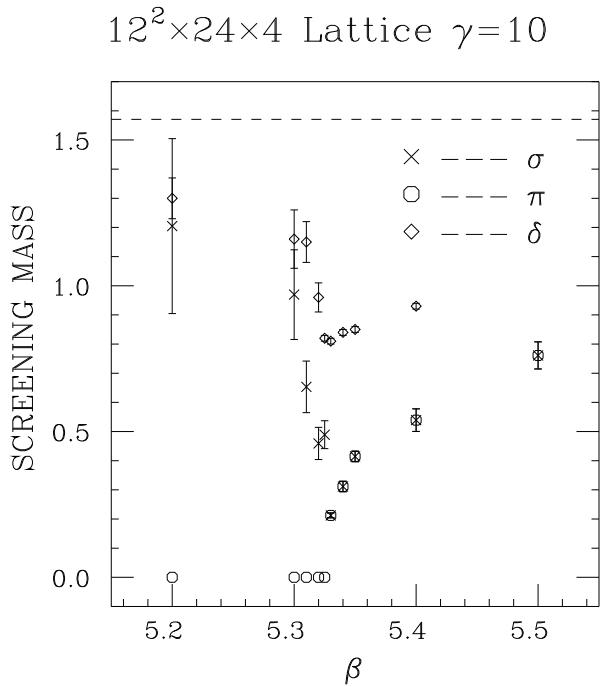}}
\caption{The $\sigma(f_0)$ and $\pi$ screening masses as functions of $\beta$.
         \label{fig:sigma-pi}}
\end{figure}

Our $\sigma$ and $\pi$ screening mass fits are shown in 
figure~\ref{fig:sigma-pi}. Below the transition the $\sigma$ mass falls
steeply as the transition is approached. What is unclear is whether it
actually falls to zero at the transition. The pion is massless in this regime.
Above the transition the $\sigma$/$\pi$ mass rises rapidly from a small and
possibly zero value at the transition, and is expected to approach 
$2\pi T = 2\pi/N_t = \pi/2$ as $\beta \rightarrow \infty$. Thus the restoration
of the $U(1) \times U(1) \subset SU(2) \times SU(2)$ chiral symmetry is 
manifest in the spectrum of screening masses/screening lengths.

In addition to the $\pi$ and $\sigma$, figure~\ref{fig:sigma-pi} shows the
screening masses in the flavor triplet scalar channel, $\delta$($a_0$). These
are obtained from measurements of the ``connected'' part of the $\sigma$
propagator (i.e. one quark ``bubble'' instead of the complete ``chain of
bubbles''). In contrast with the $\sigma$ and $\pi$, the $\delta$ always
remains heavy. We therefore have a situation where the $U_A(1)$ axial symmetry
only becomes effectively restored at temperatures well above the chiral phase
transition, in agreement with \cite{lagae,milcx}. The splitting between
$\delta$ and $\sigma$ or $\pi$ is very clear in this study because we work at
zero quark mass \footnote{Simulations closer to the continuum limit will
however be necessary in order to make sure that this splitting is not induced
by spurious symmetry breakings at moderate values of $\beta$ and $\gamma$.} The
screening mass of the $\delta$ also exhibits a jump (or rapid crossover) which
is consistent with our determination of a first order transition. 

Finally we have measured observables associated with the entropy density,
energy density and pressure. We present these as is, since tree level
extractions of entropy density, energy density and pressure are unphysical
(they require $p=\frac{1}{3}\epsilon$, the ideal gas relation, which cannot hold
through a first order transition where pressure is continuous and energy
density is discontinuous), and the one loop calculations have yet to be
calculated. In addition, for the energy density and pressure, we need zero
temperature measurements of observables at the same $\beta$ values.
The 3 observables given in table~\ref{tab:entropies} are related
to the contributions of the gauge fields, the quark fields and the chiral 
4-fermion interactions respectively, to the entropy density. 
\begin{table}
\begin{tabular}{dddd}
$\beta$    & $\beta(P_{st}-P_{ss})$ & 
$\langle\bar{\psi}D\!\!\!\!/_0\psi\rangle-\frac{3}{4}$ & 
$\frac{1}{8}N_f\gamma(\sigma^2+\pi^2)$                        \\
\hline
5.2        & 0.0001(13)   & -0.0298(5)   & 1.0362(4)          \\
5.3        & 0.0067(10)   & -0.0145(6)   & 1.0282(4)          \\
5.31       & 0.0081(11)   & -0.0125(8)   & 1.0264(4)          \\
5.32       & 0.0105(8)    & -0.0070(10)  & 1.0259(3)          \\
5.325      & 0.0116(8)    & -0.0049(11)  & 1.0250(3)          \\
5.33       & 0.0563(7)    &  0.0632(6)   & 1.0118(2)          \\
5.34       & 0.0587(6)    &  0.0672(6)   & 1.0109(4)          \\
5.35       & 0.0599(7)    &  0.0692(5)   & 1.0104(4)          \\
5.4        & 0.0605(9)    &  0.0716(4)   & 1.0094(3)          \\
5.5        & 0.0588(9)    &  0.0738(3)   & 1.0087(3)
\end{tabular}
\caption{Observables relevant to entropy density, energy density and pressure
as functions of $\beta$.\label{tab:entropies}}
\end{table}
What is most noticable is that each of these quantities shows a sizable jump
at the transition indicating that there will indeed be a significant increase
in entropy across this transition. This helps substantiate our conclusion that
the transition is first order. On the other hand, the variation of each of these
quantities on either side of this transition is modest. 

\section{Summary and conclusions}

The action for lattice QCD with 2 flavours of staggered quarks is modified
by the addition of a chiral 4-fermion interaction. The chosen 4-fermion
interaction has a $U(1) \times U(1) \subset SU(2) \times SU(2)$ symmetry,
where $SU(2) \times SU(2)$ is the normal $SU(N_f) \times SU(N_f)$ chiral
flavour symmetry of QCD with massless quarks. This irrelevant interaction
alows us to simulate at zero quark mass by rendering the Dirac operator 
non-singular.

We have performed simulations at $\gamma=3/\lambda^2=10$, a relatively weak
value for the 4-fermion coupling, and zero quark mass at both zero and finite
temperatures. The zero temperature calculation was performed on an 
$8^3 \times 24$ lattice at $\beta=6/g^2=5.4$, and was meant as a precurser to
performing serious hadron spectroscopy with this action. Our major result was
clear evidence of a massless Goldstone pion, the other hadrons having
relatively large masses (in lattice units) for gauge couplings and hence 
lattice spacings this large. The effectiveness of the 4-fermion term in
rendering the Dirac operator non-singular is reflected in that the average 
number of conjugate gradient iterations required to invert the Dirac operator
was $< 250$.

The finite temperature simulations were performed on $12^2 \times 24 \times 4$
and $8^3 \times 4$ lattices. The finite temperature transition occured at
$\beta_c=5.327(2)$, which is to be compared with $\beta_c=5.225(5)$, estimated
for the massless quark extrapolation with the standard action \cite{karsch93}.
We have seen clear evidence that the transition is a strong first order
transition. Since we believe that the actual continuum transition should be
second order (or possibly weakly first order) \cite{pisarski} which is what
simulations with the standard action suggest
\cite{milc12,htmcgc8,ukawa96,kanaya95,karsch93}, we believe this to be due to
the size of the 4-fermion interaction, enhanced by the strength of the gauge
coupling. Again the condition of the Dirac operator was well under control, it
requiring $< 400$ conjugate gradient iterations, on average, to invert this
operator throughout the range of $\beta$'s considered. We are now running at
$N_t=4$ with $\gamma=20$, and $N_t=6$ with $\gamma=10$ and $\gamma=20$.
Preliminary results at $N_t=6$ suggest a second order transition at
$\gamma=10$, which is even clearer at $\gamma=20$. It is still too early to
draw any conclusions from the ongoing runs at $N_t=4$, $\gamma=20$. The $\pi$
and $\sigma$ screening masses showed the appropriate behaviour. The Goldstone
pion mass remained zero below the transition, and increased from zero after the
transition. The $\sigma$ mass dipped as the transition was approached from
below (it is not clear whether it actually approached zero on the cold side of
the transition), and rose, apparently from zero, above the transition where it
was degenerate with the pion mass. The other screening masses remained large
across the transition. In particular the connected part of the $\sigma$
propagator, the $\delta(a_0)$ ramained large over this region. However, flavour
symmetry violation is large enough at these $\beta$ and $\gamma$ values that it
is not clear whether this is further evidence that the flavour singlet $U(1)_A$
symmetry remains broken across the transition from hadronic matter to a
quark-gluon plasma \cite{lagae,milcx}. 

If our thermodynamics simulations at larger $N_t$ and/or $\gamma$ do indicate a
second order transition, we should be able to determine the universality class
of this transition and the critical indices, enabling us to determine the
equation of state of hadronic matter. With still larger $N_t$ we should be able
to measure asymptotic scaling and determine the position, as well as nature of
the phase transition. At zero temperature we should be able to obtain the
hadronic spectrum, first at zero mass and then at the physical quark masses
{\it without} having to perform an extrapolation in quark mass. 

We are investigating how one might overcome the difficulties with simulating
with a 4-fermion term having the more physical $SU(2) \times SU(2)$ chiral
symmetry, and perhaps $U(2) \times U(2)$ chiral symmetry. In addition we will
investigate combining this improvement to the standard lattice QCD action with
the improvements of Lepage {\it et al.} \cite{lepage}, or of the ``perfect 
action'' methods \cite{hn}. 

Preliminary results of these simulations were reported at LATTICE'97
\cite{ks}.

\section{ACKNOWLEDGEMENTS}

These computations were performed on the CRAY C-90 at NERSC. This work was
supported by the U.~S. Department of Energy under contract W-31-109-ENG-38,
and the National Science Foundation under grant NSF-PHY92-00148. We would like
to thank John Sloan and Misha Stephanov for informative conversations.


\begin{thebibliography}{999999}
\bibitem{karsch4} F.~Karsch and E.~Laermann, Phys. Rev. D {\bf 50}, 6954 (1994).
\bibitem{milc6} T.~Blum {\it et al.}, Phys. Rev. D {\bf 51}, 5153 (1995);
                C.~Bernard {\it et al.}, Phys. Rev. D {\bf 55}, 6861 (1997).
\bibitem{milc12} C.~Bernard {\it et al.}, Phys. Rev. D {\bf 54}, 4585 (1996).
\bibitem{htmcgc8} S.~Gottlieb et al., Phys. Rev. D {\bf 55}, 6852 (1997).
\bibitem{ukawa96} A.~Ukawa, Nucl. Phys. B(Proc. Suppl.) {\bf 53}, 106 (1997).
\bibitem{kanaya95} K.~Kanaya, Nucl. Phys. B(Proc. Suppl.) {\bf 47}, 144 (1996).
\bibitem{karsch93} F.~Karsch, Nucl. Phys. B(Proc. Suppl.) {\bf 34}, 63 (1994).
\bibitem{gn} D.~J.~Gross and A.~Neveu, Phys. Rev. D {\bf 20}, 3235 (1974).
\bibitem{njl} Y.~Nambu and G.~Jona-Lasinio, Phys. Rev. {\bf 122}, 345 (1961).
\bibitem{kmy} K.~I.~Kondo, H.~Mino and K.~Yamawaki, Phys. Rev. D {\bf 39},2430
              (1989).
\bibitem{brower} R.~C.~Brower, Y.~Shen and C.-I.~Tan, Boston University
                 preprint BUHEP-94-3 (1994); R.~C.~Brower, K.~Orginos and
                 C.-I.~Tan, Nucl. Phys. B(Proc. Suppl.) {\bf 42} (1995).
\bibitem{sw} B.~Sheikholeslami and R.~Wohlert, Nucl. Phys. {\bf B259}, 572 
             (1985).
\bibitem{milc} T.~Blum {\it et al.}, Phys. Rev. D {\bf 55}, R1133 (1997).
\bibitem{hands1} S.~Hands, A.~Kocic and J.~B.~Kogut, Ann. Phys. {\bf 224} 
                 (1993).
\bibitem{kim} S.~Kim, A.~Kocic and J.~B.~Kogut, Nucl. Phys. {\bf B429}, 407
              (1994).
\bibitem{hands2} S.~Hands and J.~B.~Kogut, e-print hep-lat/9705038 (1997).
\bibitem{hkk} S.~Hands, S.~Kim and J.~B.~Kogut, Nucl. Phys. {\bf B442}, 364 
              (1995).
\bibitem{pisarski} R.~Pisarski and F.~Wilczek, Phys. Rev. D {\bf 29}, 338 
                   (1984).
\bibitem{lagae} J.~B.~Kogut, J.-F.~Lagae and D.~K.~Sinclair, 
                Nucl. Phys. B(Proc. Suppl.) {\bf 53}, 269 (1997).
\bibitem{milcx} C.~Bernard {\it et al.}, Nucl. Phys. B(Proc. Suppl.) {\bf 53},
                442 (1997).
\bibitem{lepage} M.~Alford {\it et al.}, Phys. Lett. {\bf B361}, 87 (1995).

\bibitem{hn} P.~Hasenfratz and F.~Niedermayer, Nucl. Phys. {\bf B414} (1994).
\bibitem{ks} J.~B.~Kogut and D.~K.~Sinclair, Nucl. Phys. B(Proc. Suppl.)
             {\bf 53}, 272 (1997).
\end{thebibliography}
\end{document}